\documentclass[aps,prl,twocolumn,showpacs,superscriptaddress]{revtex4}
\usepackage{graphicx}
\begin{document}

\title{Superconductivity mediated by the antiferromagnetic spin-wave in chalcogenide iron-base superconductors}

\author{G. M. Zhang}

\affiliation{
Department of Physics, Tsinghua University
}

\author{Z. Y. Lu}

\affiliation{
Department of Physics, Renmin University of China
}

\author{T. Xiang}
\affiliation{Institute of Physics, Chinese Academy of Sciences, Beijing 100190, China
}
\affiliation{Institute of Theoretical Physics, Chinese Academy of Sciences, P.O. Box 2735, Beijing 100190, China}

\begin{abstract}

The ground state of K$_{0.8+x}$Fe$_{1.6+y}$Se$_2$ and other iron-based selenide superconductors are doped antiferromagnetic semiconductors. There are well defined iron local moments whose energies are separated from those of conduction electrons by a large band gap in these materials. We propose that the low energy physics of this system is governed by a model Hamiltonian of interacting electrons with on-site ferromagnetic exchange interactions and inter-site superexchange interactions. We have derived the effective pairing potential of electrons under the linear spin-wave approximation and shown that the superconductivity can be driven by mediating coherent spin wave excitations in these materials. Our work provides a natural account for the coexistence of superconducting and antiferromagnetic long range orders observed by neutron scattering and other experiments.

\end{abstract}

\maketitle

The discovery of high-T$_c$ superconductivity in iron-based superconductors\cite{Hosono} has triggered a surge of interest for the investigation of unconventional superconducting pairing mechanism. Like in cuprate superconductors, the superconducting pairing in these materials is less likely to be mediated by phonons, as suggested by the local-density-approximation (LDA) calculations\cite{Paglione10} and experimental measurements\cite{Boeri08}. On the other hand, the proximity of superconductivity to an antiferromagnetic phase suggests that the magnetic fluctuation plays an important role in the understanding of pairing mechanism.

Recently a potassium intercalated FeSe superconductor, with a nominal composition K$_{0.8}$Fe$_2$Se$_2$ and $T_c\sim 30K$, and other chalcogenide iron-based superconductors, were discovered. It reveals many important features on the interplay between superconductivity and antiferromagnetism. The muon-spin relaxation measurement indicates that a superconducting order below a critical temperature $28K$ coexists microscopically with a strong antiferromagnetic long range order formed below 478K in Cs$_{0.8}$Fe$_2$Se$_{1.96}$.\cite{Shermadini11} From neutron scattering measurement, Bao et al.\cite{Bao11} also observed this kind of coexistence in potassium intercalated FeSe samples. In particular, they found that an antiferromagnetic order with a unprecedentedly large moment 3.31 $\mu_B$/Fe occurs at a record high value of $T_N = 559 K$ below an Fe vacancy ordering temperature $\sim 578 K$, and persists to the superconducting phase in K$_{0.82(2)}$Fe$_{1.626(3)}$Se$_2$. The coexistence of superconductivity with an antiferromagnetic state of large magnetic moments and high Neel temperature has also been reported in other mono-valent element intercalated chalcogenide iron-based superconductors.\cite{Yu11,Liu11}

The coexistence of antiferromagnetic and superconducting orders is in fact not a new phenomenon. It was also reported in hole doped BaFe$_2$As$_2$\cite{ChenXH}, heave-fermion superconductor\cite{heavefermion}, and electron doped cuprates\cite{cuprate}. However, in all previous reports, the ordering moment is small due to strong antiferromagnetic fluctuation. This has led to a common believe that the superconducting order is disfavored by the antiferromagnetic long range order, although superconducting pairing can be induced or enhanced by incoherent antiferromagnetic fluctuation. Corresponding to this empirical picture, a recipe for searching unconventional superconductors is to suppress long-range magnetic order by doping or pressure. The observation of large magnetic ordering moments in the superconducting state of K$_{0.8+x}$Fe$_{1.6+y}$Se$_2$ with relatively high T$_c$ is entirely a surprise. It seems to be difficult to incorporate this fact into the framework of superconductivity induced by antiferromagnetic fluctuations.

In this paper, we propose an effective low-energy Hamiltonian to  describe the interplay between the local moments of Fe ions and the itinerant electrons in K$_{0.8+x}$Fe$_{1.6+y}$Se$_2$ and other iron-based chalcogenide superconductors. From the effective pairing potential of electrons derived from the second order perturbation, we propose that these materials are antiferromagnetic spin-wave mediated superconductors.

Let us start by considering the electronic and magnetic structures of K$_{0.8+x}$Fe$_{1.6+y}$Se$_2$. From both transmission electron microscopy\cite{LiJQ} and neutron scattering\cite{Bao11} measurements, it is found that Fe vacancies form a superstructure with a unit cell of $\sqrt{5}\times \sqrt{5}\times 1$ in K$_{0.8}$Fe$_{1.6}$Se$_2$. This enlarged crystallographic unit cell is corroborated by the observation of extra phonon modes than the tetragonal ThCr$_2$Si$_2$ structure would allow by optical\cite{WangNL} and Raman\cite{raman} scattering studies. In this sense, it is more appropriate to label K$_{0.8}$Fe$_{1.6}$Se$_2$ as a 245 (K$_2$Fe$_4$Se$_5$) compound\cite{Bao245}.

\begin{figure}[t]
\includegraphics[width=6cm]{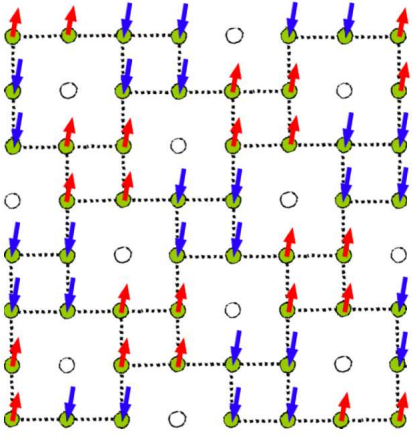}
\caption{(Color online) Schematic representation of the $\sqrt{5} \times \sqrt{5}$ superstructure of Fe vacancies in each Fe layer of  K$_{0.8}$Fe$_{1.6}$Se$_2$. Each four Fe ions on a square separated by vacanices form a cluster. These clusters form a square lattice. Fe irons within each cluster are coupled mainly by ferromagnetic exchange interactions\cite{Yan245}. But the inter-cluster interactions are antiferromagnetic. In the ground state, the clustered Fe ions form a checkerboard antiferromagnetic order. The total ordering moment of each Fe cluster is about $3.31 \times 4 \mu_B = 13.24\mu_B$.\cite{Bao11} }
\label{fig:mag}
\end{figure}

From the LDA band structure calculations\cite{Yan245}, we find that the ground state of A$_{0.8}$Fe$_{1.6}$Se$_2$ has indeed a $\sqrt{5} \times \sqrt{5}$ superstructure of Fe vacancies, and Fe moments exhibit a cluster checkboard antiferromagntic order with a moment $\sim 3.37\mu_B$, in good agreement with the neutron scattering experiment\cite{Bao11}. A schematic representation of the vacancy superstructure and the cluster checkerboard antiferromagnetic order is shown in Fig.~\ref{fig:mag}.
Different from all previous reported Fe-based superconducting materials, antiferromagnetic order in K$_{0.8}$Fe$_{1.6}$Se$_2$ occurs in a tetragonal $\sqrt{5} \times \sqrt{5}$ unit cell, maintaining the four-fold rotation symmetry. Furthermore, we find that stoichiometric A$_{0.8}$Fe$_{1.6}$Se$_2$ is an antiferromagnetic semiconductor and can be regarded as a parent compound of iron-based chalcogenide superconductors. Fe ion in this material has a 2+ valance. The cluster checkerboard antiferromagnetic order generates a self-consistent Hartree-Fock staggered field which opens a large band gap ($\sim 0.6$eV) between the conduction and valence bands.

\begin{figure}[t]
\includegraphics[width=8.8cm]{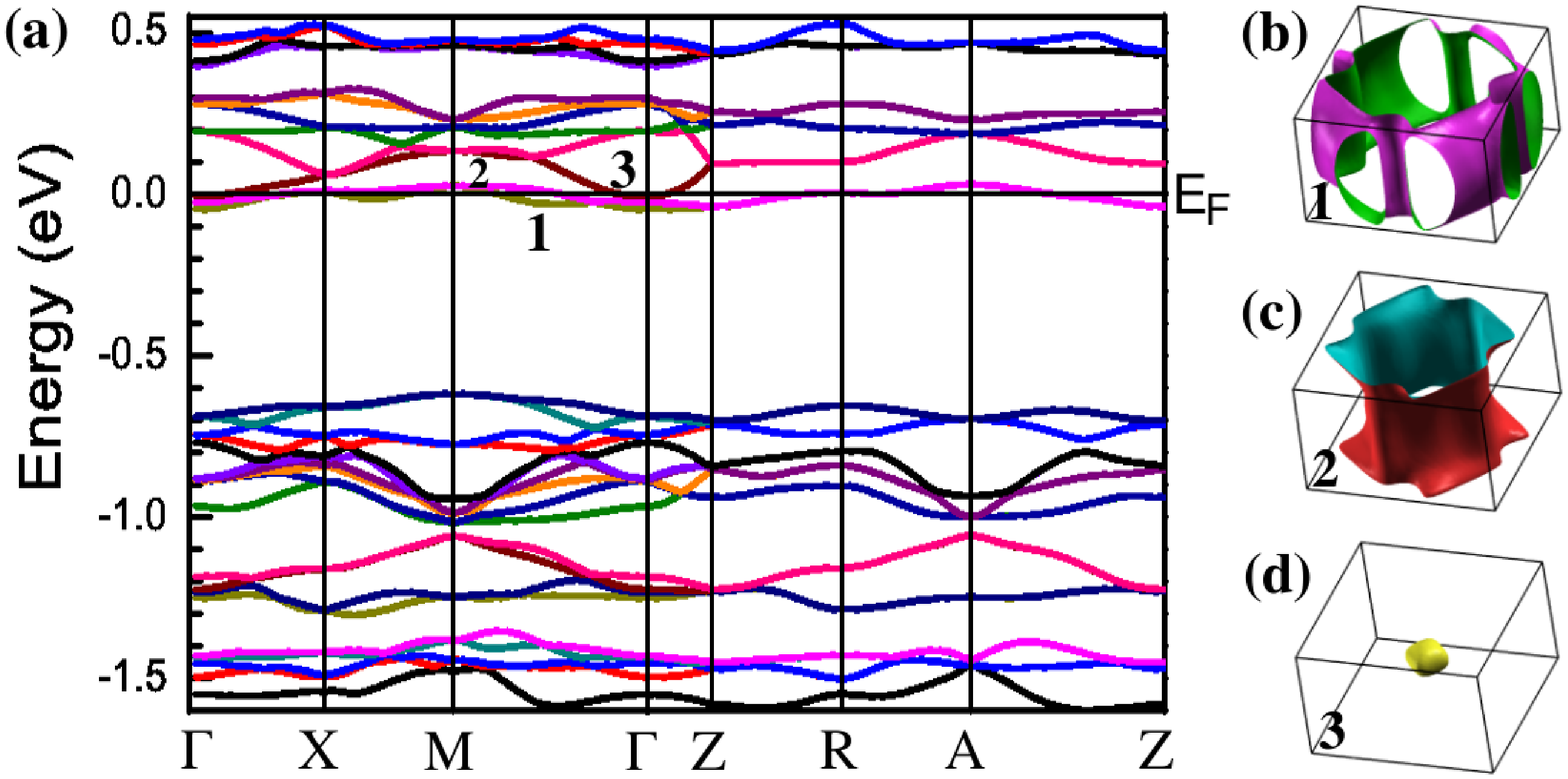}
\caption{(Color online) Band structure of KFe$_{1.6}$Se$_2$. (b-d) are the Fermi surface sheets for the three conduction bands labeled in (a). } \label{fig:band}
\end{figure}

For the stoichiometric K$_{0.8}$Fe$_{1.6}$Se$_2$, the conduction bands are empty and the valence bands are fully filled. Slightly doping electrons by increasing the K content, the conduction bands will be partially filled. But the band structure (Fig.~\ref{fig:band}) is almost unchanged. The concentration of conduction electrons is proportional to $x$ in K$_{0.8+x}$Fe$_{1.6}$Se$_2$. In most of the K$_{0.8+x}$Fe$_{1.6+y}$Se$_2$ compounds so far synthesized, the true content of K and Fe is generally less than that in the nominal composition\cite{Bao11}. This explains the optical data which show very low charge carrier concentration\cite{WangNL}, and indicates that the iron selenide superconductor is a doped antiferromagnetic semiconductor.

In K$_{0.8+x}$Fe$_{1.6+y}$Se$_2$, there are two kinds of magnetic interactions which are important to the understanding of physical properties. One is the Hund's rule coupling between the $3d$ orbitals in each Fe ion. The other is the magnetic exchange interaction between two Fe ions. Within each 4-Fe cluster shown in Fig.~\ref{fig:mag}, this exchange interaction is predominately ferromagnetic, due to a lattice distortion induced by the Fe vacancies\cite{Yan245}. But the intercluster interaction is antiferromagnetic, which is governed by the superexchange interaction mediated by Se 4p orbitals. If we denote each 4-Fe cluster as a lattice site, then an effective low energy Hamiltonian for describing physical proporties of iron chalcogenide superconductors is defined by
\begin{eqnarray}
H &=& \sum_{k\mu} \varepsilon_{\mu k} c^\dagger_{\mu k} c_{\mu k} + \sum_{\langle ij \rangle \mu \nu } J_{\mu\nu}
c^\dagger_{\mu , i}\frac{\mathbf{\sigma}}{2} c_{\mu ,i} \cdot
c^\dagger_{\nu , j}\frac{\mathbf{\sigma}}{2} c_{\nu ,j}
\nonumber\\
&& -  \sum_{i \mu \not= \nu} K_{\mu\nu} c^\dagger_{\mu , i}\frac{\mathbf{\sigma}}{2} c_{\mu ,i} \cdot c^\dagger_{\nu , i}\frac{\mathbf{\sigma}}{2} c_{\mu ,i},
\label{eq:ham1}
\end{eqnarray}
where $c_{\mu, i}=(c_{\mu ,i\uparrow}, c_{\mu, i\downarrow})$, $(\mu , \nu)$ are the band indices and $k$ is the momentum of electron. $\langle ij \rangle$ means that $i$ is a nearest neighbor of $j$. The second term describes the effective antiferromagnetic interaction between the orbitals on the neighboring two clusters (or sites). The third term describes the ferromagnetic interactions between the orbitals at site $i$. It includes the Hund's rule coupling and the ferromagnetic exchange interactions within a cluster. Similar Hamiltonian has been proposed for describing other iron-based superconductors.\cite{LuZY122}

As above mentioned, the conduction and valence bands are well separated by the energy gap in K$_{0.8+x}$Fe$_{1.6+y}$Se$_2$. In the antiferromagnetic state, the valence bands describes essentially the collective behavior of local moments. It is the superexchange interaction between these local moments that lead to the checkerboard antiferromagnetic long range order, as observed by experiments. Thus in these materials, there are well defined local moments, which are separated from the itinerant electrons in the conduction bands by the band gap. This is different from other iron-based pnictide or chalcogenide superconductors where there is no energy gap to separate local moments from itinerant electrons.

In the low energy limit, the charge fluctuation of the valence bands are frozen by the band gap. Thus only the spin dynamics of the valence bands needs to be considered. In this case, the model defined by Eq.~(\ref{eq:ham1}) can be simplified as
\begin{eqnarray}
H & = & \sum_{k\alpha }\varepsilon_{\alpha k} c^\dagger_{\alpha k} c_{\alpha k} + J \sum_{\langle ij \rangle  } \mathbf{S}_{i} \cdot \mathbf{S}_{j}
\nonumber \\
&& + \sum_{ij \alpha }\left( J^\prime \delta_{\langle ij \rangle} - K\delta_{i,j}\right) c^\dagger_{\alpha , i}\frac{\mathbf{\sigma}}{2} c_{\alpha ,i} \cdot \mathbf{S}_{j},
\label{eq:ham2}
\end{eqnarray}
where $\alpha$ stands for the conduction bands and $\mathbf{S}_i$ is the local spin at site $i$. In obtaining this expression, we have ignored the on-site ferromagnetic interaction and the inter-site superexchange interaction terms between two conduction elections. For simplicity, we have assumed that the ferromagnetic coupling constants and the antiferromagnetic superexchange constants between a conduction election and a local spin do not depend on the orbital index $\alpha$, defined by $K$ and $J^\prime$, respectively.

If holes are doped to K$_{0.8}$Fe$_{1.6}$Se$_2$, the Fermi level falls into the valence bands and the charge current is conducted by holes. In the limit of low hole density, the effective Hamiltonian is still given by Eq.~(\ref{eq:ham2}), but both $K$ and $J^\prime$ change sign, namely, the $K$-term becomes antiferromagnetic and the $J^\prime$-term becomes ferromagnetic. Unlike in the electron doped case, the onsite Kondo type of interaction is a relevant perturbation. It has a strong screening effect on the local moments. Thus we believe that the phase diagram of K$_{0.8+x}$Fe$_{1.6}$Se$_2$ is highly particle-hole asymmetric.

For other iron-based superconductors, all 3d bands of Fe are strongly mixed and there is not a gap to suppress the charge fluctuation between local moments and conducting electrons. In this case, it is difficult to give a clear definition of local moments. Nevertheless, it is believed that physical properties can still be qualitatively understood from an effective Hamiltonian similar to Eq.~(\ref{eq:ham2}), which includes the on-site Hund's rule coupling and the inter-site antiferromagnetic superexchange interaction\cite{WengZY,KuWei}.

For K$_{0.8}$Fe$_{1.6}$Se$_2$, the total magnetic moment of each cluster is close to $3.31\times 4 \mu_B = 13.24\mu_B$ according to the neutron scattering measurement\cite{Bao11}. The corresponding spin value should therefore be larger than 6.5 if its $g$-factor is 2. For such a large spin system, the $J$-term in Eq.~(\ref{eq:ham2}) can be well treated by the linear spin-wave approximation.

In order to examine the contribution of the ferromagnetic coupling term in Eq~(\ref{eq:ham2}) to the superconducting pairing, let us first omit the $J^\prime$ term. In the antiferromagnetic state, the on-site electron-spin interaction term can be separated into the longitudinal and transverse parts. The former provides a staggered field to the itinerant electrons, and the latter gives rise to the coupling between electrons and spin wave excitations. If we ignore the interacting terms involving two-magnon excitations, then the Hamiltonian under the linear spin wave approximation can be written as
\begin{eqnarray}
H_{0} &=&\sum_{k\sigma}
D_{k\sigma }^{\dagger }\left(
\begin{array}{cc}
\xi^+_{k} & 0 \\
0 & \xi^-_{k}
\end{array}
\right) D_{k\sigma }
\nonumber \\
&& +\sum_{k}\omega _{k}\left( \alpha _{k}^{\dagger }\alpha
_{k}+\beta _{k}^{\dagger }\beta _{k}\right) ,
\\
H_1 & = & \eta \sum_{kq} D_{k+q\downarrow }^{\dagger
}W_{k+q} B_{q} W_{k} D_{k\uparrow }+h.c.  \label{eq:H1}
\end{eqnarray}
For the conciseness, we have ignored the band index $\alpha$ in the above expressions. In $H_0$, $\alpha_k$ and $\beta_k$ are the magnon operators, $\omega _{k} = JS\sqrt{4-\gamma _{k}^{2}} $ is the energy dispersion of spin wave excitations, $\gamma _{k} =\cos k_{x}+\cos k_{y}$.
\[
\xi^\pm_{k}=\frac12 \left[ \varepsilon_{k}+\varepsilon_{k+Q}\pm \sqrt{\left( \varepsilon_{k}-\varepsilon _{k+Q}\right)^{2}+\left( K S\right) ^{2}} \right]
\]
are the modified band dispersions in a staggered field. $D_{k \sigma}^\dagger = (c_{k ,\sigma}^\dagger, c_{k+Q ,\sigma}^\dagger ) W_{k} $ are the electron operators of these bands, $Q=(\pi, \pi)$, and
\[
W_{k } = \left(\begin{array}{cc}
\cos \theta _{k} & \sin \theta _{k} \\
-\sin \theta _{k} & \cos \theta _{k}
\end{array} \right)
\]
is the corresponding unitary transform matrix.
In $H_1$, $\eta  = -K\sqrt{S}/(2\sqrt{2})$ and
\[
B_{q} =\left( v_{q}+u_{q}\right) \left( \beta _{q}^{\dagger }+\alpha_{q}\right) +\left( u_{q}-v_{q}\right) \left( \beta_{q}^{\dagger }-\alpha_{q}\right) \sigma _{1}.
\]
$u_{k} =\sqrt{\frac{1}{2}+\frac{JS}{2\omega _{k}}} $ and
$v_{k} = -\mathrm{sgn}\left( \gamma _{k}\right) \sqrt{-\frac{1}{2} +\frac{JS}{2\omega_{k}}} $.

\begin{figure}[t]
\includegraphics[width=8cm]{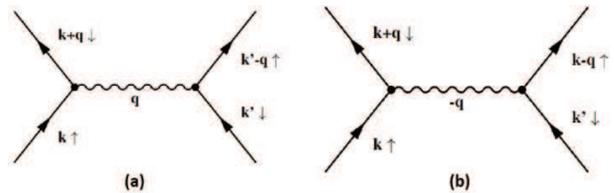}
\caption{Interaction between two electrons via a single spin-wave exchange. } \label{fig:feymann}
\end{figure}

The effective interaction between electrons can be obtained by integrating out all magnon operators.  To the second-order perturbation in $H_1$ (which is schematically represented by the Feymann diagram Fig.~\ref{fig:feymann})\cite{Frohlich}, we find that the effective electron-electron interacting potential mediated by spin wave excitations is given by
\begin{equation}
V =\sum_{qkk'}
D_{k^\prime +q\downarrow }^{\dagger } D_{k\uparrow }^{\dagger } V_{kk^\prime q} D_{k^\prime \uparrow } D_{k+q\downarrow }
\end{equation}
where
\[
V_{kk^\prime q}
=\eta^{2}\sum_{\sigma =\pm }\frac{
\sigma A_{k^{\prime} q \sigma } A_{k q \sigma }^\dagger
}{\xi _{k+q}-\xi _{k}+\sigma \Omega _{q}}
\]
and
\[
A_{k,q,\pm } = \left( v_{q}+u_{q}\right) W_{k+q}W_{k}\pm \left(
u_{q}-v_{q}\right) W_{k+q}\sigma _{1}W_{k}
\]
This potential does not possess the SU(2) spin rotation symmetry because this symmetry is broken in the antiferromagnetic long range ordered state.

In the Cooper channel, the interacting potential can be simplified as
\begin{equation}
V^{co} = \eta ^{2}\sum_{kk'}\frac{1}{\sqrt{4-\gamma_k^2} }
\frac{2 g_{kk'} \omega _{k+k^{\prime }}}{ \left( \xi _{k^{\prime }}-\xi _{k}\right)^{2}-\omega _{k+k^{\prime }}^{2} }
\end{equation}
where $g_{kk'}  = (2+\gamma _{k}) \sin^2( \theta_{k} - \theta_{k'} ) - (2-\gamma _{k}) \cos^2 (\theta_{k} + \theta_{k'})$. As $g_{kk'}$ is positive when $\gamma_k$ is close to 2, one can show that the effective potential $V^{co}$ is attractive in certain $k$-space around the Fermi surface. Similar to the electron-phonon case, this attractive interaction can drive the system into a superconducting state. Furthermore, it can be shown that the dominant channel of intraband potential has an even parity. Thus the superconducting pairing happens mainly in the singlet channel, consistent with NMR experiments\cite{Yu_NMR}. The pairing symmetry depends on the detail of the band structure, especially the structure of Fermi surfaces. In this multi-band system, it also depends on the inter-band scattering potential.

It is helpful to make a comparison with conventional phonon mediated superconductors. In a phonon mediated superconductor, there are two characteristic energy scales, the Debye frequency $\omega_D$ and the electron-phonon coupling $\lambda$. In a spin-wave mediated superconductor, these energy scales are proportional to the antiferromagnetic superexchange coupling constant $JS$ and the on-site ferromagnetic coupling constant $K\sqrt{S}$, respectively. Both $JS$ and $K\sqrt{S}$ in these materials could be quite large, this may explain why T$_c$ is so high.

The electron-spin interaction defined by the $J^\prime$ term in Eq.~(\ref{eq:ham2}) can be similarly treated, only the $B_q$ matrix defined in Eq.~(\ref{eq:H1}) needs to be modified to include the geometric effect of interaction. Under the linear spin-wave approximation, this term is similar to the effective electron-spin-wave interaction that was used for studying the spin-wave mediated superconductivity in high-T$_c$ cuprates in Ref.~[\onlinecite{Hida,Sushkov}]. As discussed in these references, this kind of interaction can also drive the system into a superconducting state. Thus the coexistence of superconducting and antiferromagnetic long range order is a universal feature of the model Hamiltonian, defined by Eq.~(\ref{eq:ham2}).

To summarize, we propose a minimal model to describe low-energy physical properties of iron-based chalcogenide superconductors. From the pairing potential of electrons derived from the second order perturbation, we suggest these materials to be antiferromagnetic spin wave mediated superconductors. This result gives a natural account for the experimental observation on the coexistence of superconducting and antiferromagnetic long range orders with large magnetic moments. K$_{0.8+x}$Fe$_{1.6+y}$Se$_2$ and related materials presents a unique and ideal limit to investigate unconventional mechanism of superconductivity.
Further studies on the pairing symmetry and electromagnetic response functions in these materials may help us to resolve many disputable issues.

We would like to thank W. Bao, D.H. Lee, and N.L. Wang for helpful discussions. This work is supported by NSFC and the grants
of National Program for Basic Research of MOST of China.

\end{document}